\documentclass[12pt,a4paper]{article}

\usepackage{amsfonts}
\usepackage{amsmath}

\newcommand{\danger}[1]{\textbf{#1}}

\input{epsf}             
\def\epsfcenter#1{{\vcenter{\hbox{\epsfbox{#1}}}}} 

\usepackage[dvips]{graphicx}

\newtheorem{theorem}{Theorem}

\newtheorem{definition}{Definition}

\numberwithin{myeqn}{section}

\begin{document}

\title{\danger{Observables in 3-dimensional quantum gravity 
and topological invariants}}
\author{\danger{J. Manuel Garc\'\i a-Islas}
\footnote{This paper is based on a part of the Ph.D thesis of the author
at the University of Nottingham, UK \cite{jm}. 
The work was motivated by a joint colaboration with
John W.Barrett \cite{b-jm}.}}

\date{}

\maketitle

\hspace{3.1cm}Centro de Investigacion en Matem\'aticas

\hspace{5.1cm} A.P. 402, 36000

\hspace{4.2cm} Guanajuato, Gto, M\'exico

\hspace{4.5cm} email: islas@cimat.mx

\vspace{1.3cm}

\danger{Abstract}. In this paper we report some results on the expectation values 
of a set of observales introduced for 3-dimensional Riemannian quantum gravity 
with positive cosmological constant,
that is, observables in the Turaev-Viro model.
Instead of giving a formal description of the obsevables, we just formulate 
the paper by examples. This means that we just show how an idea
works with particular cases and give a way to compute
'expectation values' in general by a topological procedure.   

\section{Introduction}

The definition of good observables for quantum gravity is one of the most impotant problems.
In this paper we introduce a set of observables in the Turaev-Viro model of 3- dimensional quantum gravity
with positive cosmological constant. We also describe a very
natural way to define their expectation values. 
Instead of giving a very formal and rigourous description
of these set of observables and of their expectation values we just show how an idea
works for some particular examples. The interesting thing is that their expectation value
is related to topological invariants of such observables. 
These observables are thought as graphs, knots or links 
embedded in a 3-dimensional manifold. When the manifold is $S^{3}$, the examples
show us a way to compute this topological invariant for general cases.

We divide this paper as follows: In section
2 we briefly recall the Turaev-Viro model as a spin foam model of 3-dimensional 
Riemannian quantum gravity with positive cosmological constant. In section 3 we introduce the
notion of observables in this model as well as their expectation value. 
In section 4 we apply the introduced concepts of section 3
in order to compute the expectation value of some particular examples.
In section 5 we formulate an idea of general observables in the framework of
chain-mail decompositions which leads us to   
compute a topological invariant of knots. We refer to this topological invariant
as the chaim-mail expectation value of a knotted observable. Moreover, the idea
is considered for the prticular case of knots embedded in $S^{3}$,
We also show its topological invariance as well as
give a explicit value of the cahin-mail expectation value of a general knot or link in $S^{3}$.
Finally in section 6 we give the conclusion and a proposal of future work on the subject.

\section{The Turaev-Viro model as a spin foam model of quantum gravity}

The Turaev-Viro partition fuction is an improved
regularization of the Ponzano-Regge model \cite{pr}. 
It is a spin foam model of Riemannian quantum gravity with
positive cosmological constant.
 Moreover, the Turaev-Viro sum gives topological invariants of 3-dimensional manifolds
$M$.

Let $M$ be a closed, oriented three dimensional manifold(we can say it is our
three dimensional space-time). 
Take a triangulation $\triangle$ of $M$ with $n_0$ 
vertices, $n_1$ edges $e_i$, $n_2$ faces $f_j$ and $n_3$ tetrahedra $t_k$. 

We construct a spin foam model for 3-dimensional Riemannian
quantum gravity with positive cosmological constant by using the dual
complex $\mathcal{J}_{\Delta}$ to our
triangulation $\Delta$ of the manifold.
\footnote{The description
of the Turaev-Viro model by using the triangulation of the manifold is
an equivalent one and it is the one we will use for the description of our observables.
The triangulation description is found in \cite{tv}.}
This spin foam construction which uses the dual complex can be found in
\cite{kl}.

In order to construct the model we label each face of the complex
$\mathcal{J}_{\Delta}$ by an 
irreducible representation of the quantum group
$SU(2)_{q}$. 
The model has an integer parameter $r\ge3$, from which we define the root of
unity
$q=A^{2}=e^{ \frac{i\pi}{r}}$. 
Since $q$ is a
root of unity the number of irreducible representations is finite. A
representation can be indexed by an non-negative half-integer $j$, the spin,
from the set $L=\{0,1/2,1,\ldots, (r-2)/2 \}$.

We define a state as a map from the set of faces of the dual complex 
$\mathcal{J}_{\Delta}$, to the set $L$. 
A state is called admissible if at each edge of the
complex $\mathcal{J}_{\Delta}$, the labels $(i,j,k)$ of the three faces that are
adjacent to the edge satisfy

\[ 0 \leq i , j , k \leq \frac{r-2}{2} \]

\[ i \leq j+k , \  j \leq i+k , \  k \leq i+j \]

\[ i+j+k \equiv mod \ 1 \]

\[ i+j+k \leq (r-2) \]

The state sum model is then given by

\begin{equation}
Z(M)= N^{-n_{0}} \sum_{S} \prod_{f} A(f) \prod_{e} A(e)
\prod_{v} A(v)
\end{equation}
where the sum is carried over the set of all admissible states $S$
and
$A(f), A(e), A(v)$ are the face, edge, and vertex amplitudes
respectively and $N$ is a normalisation factor which we describe below.

These amplitudes are given by the evaluation of spin networks such as

\[ A(f)= \epsfcenter{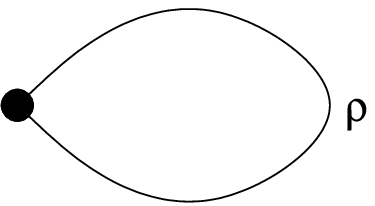} \]

\[ A(e)= \frac{1}{\epsfcenter{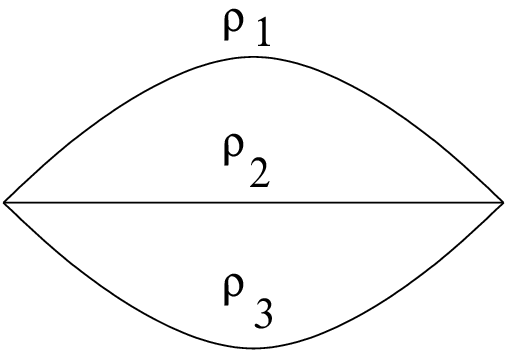}} \]

\[ A(v)= \epsfcenter{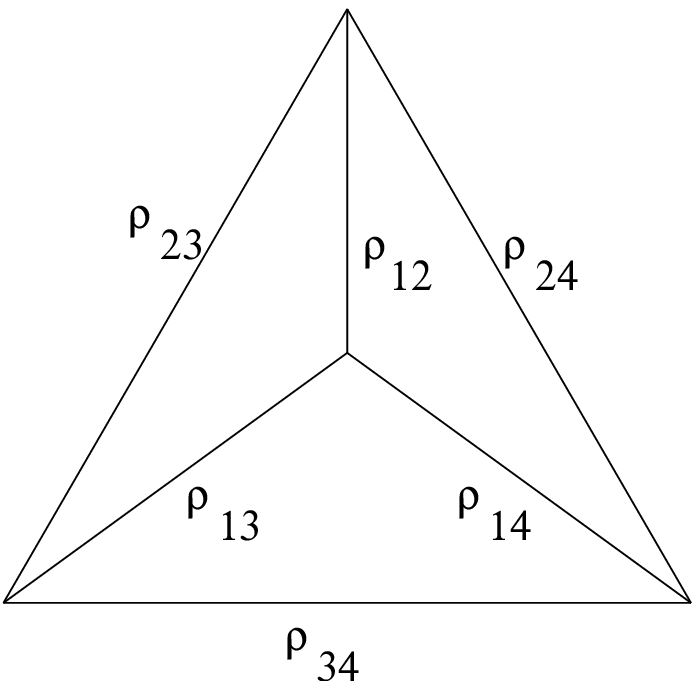} \]

The evaluation of such spin networks is given by the Kauffman bracket of the 
respective graph \cite{kl}.

For instance, the amplitude given to the faces is given by
the quantum dimension of the representation, which is given by the formula 
$\dim_{q}(j)=(-1)^{2j}[2j+1]_{q}$, where $[n]_{q}$ is the quantum number
$$[n]_{q}= \frac{q^{n}-q^{-n}}{q-q^{-1}}$$
The evaluation of the tetrahedron combined with the evaluation
of theta symbols gives the so called quantum $6j$-symbol. 

The normalisation factor is given by $N=\sum_{l\in L} \dim_q(l)^2$.

The state sum model is independent from the triangulation of the given manifold 
$M$. In terms of the dual complex we can pass from one dual complex to another
one by a finite sequence of moves known as Matveev-Piergallini  \cite{kl}.
  
However the triangulation transformations are simpler and 
Pachner shows in \cite{pa} that any two triangulations of a given
3-dimensional manifold $M$ are related by a sequence of the 
moves of figure 1 and figure 2.

\begin{figure}[h]
\begin{center}
\includegraphics[width=0.5\textwidth]{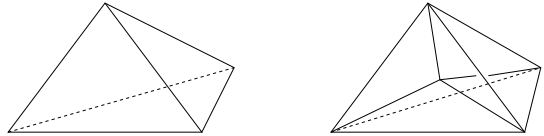}
\caption{1-4 Pachner move}
\end{center}
\end{figure}

\begin{figure}[h]
\begin{center}
\includegraphics[width=0.5\textwidth]{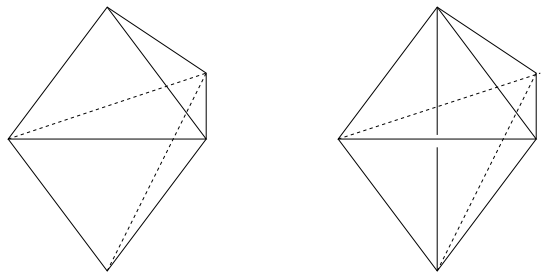}
\caption{2-3 Pachner move}
\end{center}
\end{figure}
The partition function $Z(M)$ is then invariant under the Pachner moves being then
an invariant of the 3-dimensional space-time manifold $M$.
It is not difficult to prove that the Turaev-Viro partition function is
invariant under the two Pachner moves. The second one follows easily from the 
Biedenharn-Elliot identity, and the first one follows by a direct calculation.

\section{Observables in the Turaev-Viro spin foam model}

We now introduce our observables, define its expectation value 
and compute the 
expectation value of some particular examples. 
Our observables are related to graphs contained in the
triangulated 3-dimensional manifold $M$, but these observables are part of the
triangulation itself. 
This will give us invariants of graphs in a 3-dimensional
manifold. 

A study of different observables for the Turaev-Viro model
from ours can be found in  
\cite{aw}, \cite{bd}, \cite{ks}.

We will consider the triangulation description of the
Turaev-Viro model. 
\footnote{Now we describe our observables by using the triangulation of
our 3-dimensional manifold instead of the dual domplex. The idea works
anyway in both contexts.}
Let $M$ be our triangulated 3-dimensional compact space-time manifold

\begin{definition}
We define our observables $\mathcal{O}$ to be any subset of interior edges
of our triangulated manifold $M$. We denote them by 
$\mathcal{O}=(e_{1}, e_{2},...,e_{n})$. 
If any of these edges $e_{i}$ for $i=1,2,...,n$, intersects the boundary
$\partial(M)$, that is, $\partial(M) \cap e_{i} \neq \emptyset$, 
then $\partial(M) \cap e_{i}=v$ for $v$ a vertex of the triangulation.    
\end{definition}

Given our observable subset $\mathcal{O}$, we label its edges by 
irreducible representations of our quantum group $SU(2)_{q}$. 
If $\mathcal{O}=(e_{1}, e_{2},...,e_{n})$ is our observable with edges
$e_{1}, e_{2},...,e_{n}$, we
denote the labelling of its edges by $j_{1}, j_{2},...,j_{n}$ respectively.

Consider now the same partition function of Turaev-Viro with the only 
difference that we sum over all admissible states for our triangulation
$\triangle$ 
except that now we keep the labelling of our observable edges fixed.   

We denote this sum as
 
\begin{equation} 
Z(M,\triangle,\mathcal{O})[j_1,j_2,\ldots,j_n]= 
\sum_{S|\mathcal{O}} \prod_{f} A(f) \prod_{e} A(e)
\prod_{v} A(v)
\end{equation}
It is clear that the function (2) is a function of the fixed variables
$j_1,j_2,\ldots,j_n$, as well as of the boundary fixed variables.
We ignore the boundary fixed variables and concentrate only on the observable
fixed variables. 

We have that,

\begin{equation}
\sum_{j_1\in L} \sum_{j_2\in L}\ldots \sum_{j_n\in L}  
Z(M,\triangle,\mathcal{O}) [j_1,j_2,\ldots,j_n]=Z(M,\triangle).
\label{sumequation} 
\end{equation}
We state the results in terms of a
`vacuum expectation value'

\begin{equation}
W(M,\triangle, \mathcal{O}) [j_1,j_2,\ldots,j_n]=
\frac{Z(M, \triangle,\mathcal{O}) [j_1,j_2,\ldots,j_n]}{Z(M,\triangle)}
\end{equation}
This sums to 1.
The $W$ are of course only defined when $Z\ne0$.
The  observable expectation value depends only on the subset of edges
and on the representations fixed on its edges. This gives us a way
to think of this expectation value as a topological invariant of our graph observable
$\mathcal{O}$. Their physical interpretation is shown with a particular example,
and in section 4, a relation to some other field theories such as conformal field theory
is found.

Let us consider some examples and see how this idea works. These examples are
called the one edge observable, the triangle observable and the
square observable.

We will show how the computation of the expectation value works with the one
edge observable and just give the final result for the other two cases.

\subsection{Examples}

\danger{The one edge observable}

\bigskip

Let $\mathcal{O}$ consists of a single edge $e$ which may intersect the boundary
in a single vertex only. That is, at least one vertex of $e$ lies
in the interior of $M$. For this case we prove that 

\begin{equation}
W(M, \mathcal{O})[j]= \frac{1}{N} dim_{q}(j)^{2} 
\end{equation}
where $N$ is the constant defined previously as $N= \sum_{i} dim_{q}(i)^{2}$.

Note first that clearly these numbers sum to one when we vary $j$, and are
clearly positive. There is then a natural physical interpretation of the above
expectation value as the probability that a physical quantity takes the value
$j$ \cite{b}.

Now, the link of our edge observable is a ball $B^{3}$. Let us suppose first  
that $\mathcal{O}$ is
inside a tetrahedron, as in the figure 3.

\begin{figure}[h]
\begin{center}
\includegraphics[width=0.3\textwidth]{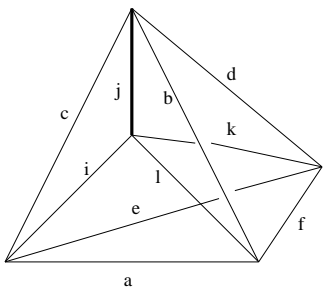}
\caption{Edge observable inside a tetrahedron}
\end{center}
\end{figure}
Now, cut this tetrahedron out of the 3-dimensional manifold and consider it as a
3-dimensional manifold with boundary. Actually it is the $B^{3}$ ball.
Take the partition function of this ball $B^{3}$, keeping in mind that it will
be a function of the fixed labelled boundary, and of the fixed labelled one edge
observable $\mathcal{O}$. 
Using the Turaev-Viro partition function with boundary we then have 
  
\[ Z(B^{3})[j]= N^{-3} dim_{q}(a)^{1/2} \cdots dim_{q}(f)^{1/2} \sum_{i,k,l} 
dim_{q}(i) dim_{q}(k) dim_{q}(l) dim_{q}(j) \times \]
\[ \qquad \qquad \times 
\left|
\begin{matrix} a&b&c\\j&i&l\end{matrix} 
\right|_{q}
\left|
\begin{matrix} f&e&a\\i&l&k\end{matrix} 
\right|_{q}
\left|
\begin{matrix} d&b&f\\l&k&j\end{matrix} 
\right|_{q} 
\left|
\begin{matrix} c&e&d\\k&j&i\end{matrix} 
\right|_{q} \]
Note that $j$ is being kept fixed.

Summing over $k$ and using the Biedenharn-Elliot identity and the symmetries of 
the $6j$ symbol we have that\footnote{The Biedenharn-Elliot identity and the symmetries
of the $6j$-symbol can be found in \cite{kl}} 

\[ \sum_{k} dim_{q}(k) \left|
\begin{matrix} f&e&a\\i&l&k\end{matrix} 
\right|_{q} 
\left|
\begin{matrix} d&b&f\\l&k&j\end{matrix} 
\right|_{q}\left|
\begin{matrix} c&e&d\\k&j&i\end{matrix} 
\right|_{q} 
= \sum_{k} dim_{q}(k) \left|
\begin{matrix} a&f&e\\k&i&l\end{matrix} 
\right|_{q} 
\left|
\begin{matrix} c&e&d\\k&j&i\end{matrix} 
\right|_{q}
\left|
\begin{matrix} b&d&f\\k&l&j\end{matrix} 
\right|_{q} \]

\[ = \left|
\begin{matrix} a&c&b\\d&f&e\end{matrix} 
\right|_{q}\left|
\begin{matrix} a&c&b\\j&l&i\end{matrix} 
\right|_{q} 
= \left|
\begin{matrix} a&b&c\\d&e&f\end{matrix} 
\right|_{q}\left|
\begin{matrix} a&b&c\\j&i&l\end{matrix} 
\right|_{q} \]
So we can write now

\[ Z(B^{3})[j]= N^{-3} dim_{q}(a)^{1/2} \cdots dim_{q}(f)^{1/2} 
\left|
\begin{matrix} a&b&c\\d&e&f\end{matrix} 
\right|_{q} \]

\[ \qquad \qquad \times \sum_{i,l} 
dim_{q}(i) dim_{q}(l) dim_{q}(j)
\left|
\begin{matrix} a&b&c\\j&i&l\end{matrix} 
\right|_{q}^{2} \]
Summing over $l$ and using the orthogonality and symmetry propierties gives

\[ \sum_{l} dim_{q}(l)
\left|
\begin{matrix} a&b&c\\j&i&l\end{matrix} 
\right|_{q} \left|
\begin{matrix} a&b&c\\j&i&l\end{matrix} 
\right|_{q} 
= \sum_{l} dim_{q}(l)
\left|
\begin{matrix} a&i&l\\j&b&c\end{matrix} 
\right|_{q} \left|
\begin{matrix} a&i&l\\j&b&c\end{matrix} 
\right|_{q} = \frac{1}{dim_{q}(c)} \]
So that 

\[ Z(B^{3})[j]= N^{-3} 
\dim_{q}(a)^{1/2} \cdots dim_{q}(f)^{1/2} 
\left|
\begin{matrix} a&b&c\\d&e&f\end{matrix} 
\right|_{q} \]

\[ \qquad \qquad \times
\sum_{i} \frac{\delta_{ijc}}{dim_{q}(c)} 
dim_{q}(i) dim_{q}(j) \]
where $\delta_{ijc}=1$ if $(i,j,c)$ are admissible and $0$ otherwise.

Taking the sum over $i$

\[ \sum_{i} \delta_{ijc} dim_{q}(i)= dim_{q}(c) dim_{q}(j) \]
finally gives

\[ Z(B^{3})[j]= N^{-3} dim_{q}(a)^{1/2} \cdots dim_{q}(f)^{1/2}
\left|
\begin{matrix} a&b&c\\d&e&f\end{matrix} 
\right|_{q} dim_{q}(j)^{2} \]
Now glue the ball $B^{3}$ back into $M$, so that

\[ Z(M, \mathcal{O})[j]= dim_{q}(j)^{2} \sum_{S,a,b, \cdots f}
dim_{q}(a) 
\cdots dim_{q}(f) \prod_{edges} dim_{q}(i) \]

\[ \qquad \qquad \times
\left|
\begin{matrix} a&b&c\\d&e&f\end{matrix} 
\right|_{q} \prod_{tetrahedra} \left|
\begin{matrix} a'&b'&c'\\d'&e'&f'
\end{matrix} 
\right|_{q} \]
where we have a product of the $6j$ symbol assigned to the tetrahedron 
$\{ \{ a,b,c \}, \{ b,d,f \},$ 

$\{ c,d,e \}, \{ a,e,f \} \}$ with the $6j$ symbols
assigned to all the remaining tetrahedra of the 3-manifold. 

As we have that

\[ W(M, \mathcal{O})[j]= \frac{Z(M, \mathcal{O})[j]}{Z(M)} \]
and in our sum $Z(M, \mathcal{O})[j]$ we have one less vertex than in $Z(M)$, we
have that

\[ W(M, \mathcal{O})[j]= \frac{1}{N}
dim_{q}(j)^{2} \]
as desired.

It can be seen that the expectation value does not depend on how the edge observable
lives inside the triangulated manifold. In particular, the link of a vertex can be given
by a more complex polyhedron which is just a triangulated 3-ball.(see figure 4) 

\begin{figure}[h]
\begin{center}
\includegraphics[width=0.4\textwidth]{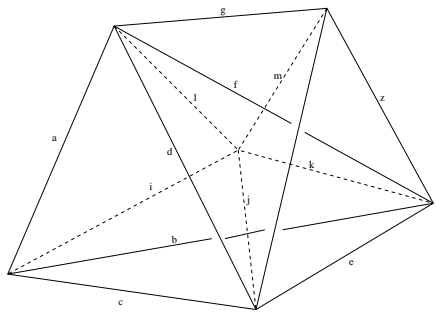}
\caption{One edge observable inside a polyhedron}
\end{center}
\end{figure}

\newpage

\danger{The triangle observable}

\bigskip

We now consider a triangle observable.\footnote{The computations can be found in \cite{jm}}
Suppose the manifold $M$ contains a triangle(see figure 5), whose edges form $\mathcal{O}$, 
labelled by $i$, $j$ and $c$. 
Let $N_{i,j,c}$ be the dimension of the space of intertwiners, 
i.e. equal to 1
if the spins are admissible and 0 otherwise. We have that

\begin{figure}[h]
\begin{center}
\includegraphics[width=0.3\textwidth]{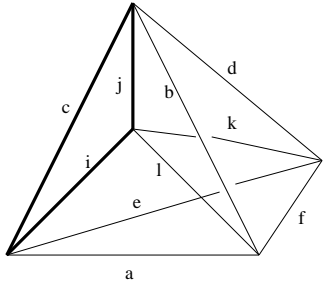}
\caption{Triangle observable inside a tetrahedron}
\end{center}
\end{figure}

\begin{equation} 
W(M, \mathcal{O})[i,j,c]= 
\frac{1}{N^2} \dim_{q}(i) \dim_{q}(j) \dim_{q}(c) N_{i,j,c} 
\end{equation}

\newpage

\danger{The square observable}

\bigskip

Now suppose the manifold $M$ contains a square observable $\mathcal{O}$ whose
edges are labelled by the representations $i, j, m, n$. Then we prove that

\begin{figure}[h]
\begin{center}
\includegraphics[width=0.5\textwidth]{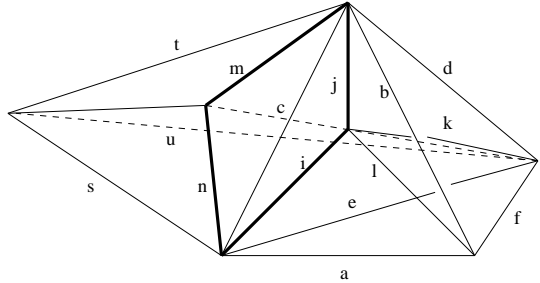}
\caption{A square observable}
\end{center}
\end{figure}

\begin{equation} 
W(M, \mathcal{O})[i,j,m,n]= 
\frac{1}{N^{3}} dim_{q}(i)dim_{q}(j)dim_{q}(m)dim_{q}(n) N_{i,j,c} N_{m,n,c}
\end{equation}

In this section we have shown by examples how to calculate the expectation value of some
our observables. This idea may be extended to more complicated observables
and the computational methods are similar. It is really easy to compute the expectation
values of observables given by trees, not knotted cycles of any length,
and any combination of these examples.

Moreover, we notice that the expectation value of these
simple examples do not depend on the manifold in which they are embedded
but only in the representations which label them.  
In general it is the same for trees and not knotted cycles. But what happens if our
observable is knotted, or it is link. We think that the above triangulation methods
cannot be applied and we requiere of a more sophisticated method. This is done by
the chaim-mail method \cite{r}. In the next section we describe a way to deal
with knotted observables in the case in which the observable is embedded in $S^{3}$.

\section{Knotted observables in $S^{3}$ and their chain-mail expectation value}

We now describe the idea of knotted observables in $S^{3}$.
Following the same strategy of the paper, we just describe
the idea a bit informally, but we work on examples to see
the way it works.
 
We define a way to compute their chain-mail expectation value by
introducing this picture. 
The chain-mail expectation value that is described by using this chain-mail
picture is analogous to the expectation value of our observables
of the previous chapter and it is restricted to 
knots and links living in $S^{3}$. 
A formal description of the relationship of
the expectation value of the previous section and of the
chain-mail one will appear in \cite{b-jm}. 

In this section we restrict ourselves and only describe the
way to obtain the chain-mail picture of our observable knot or link.
We describe how to compute their chaim-mail expectation value and then
we prove that it is a topological invariant by proving that it remains
invariant under the Reidemeister moves. 
Finally, we give a general formula for the expectation value of any 
knot or link observable.

Consider a knot or link

\begin{figure}[h]
\begin{center}
\includegraphics[width=0.3\textwidth]{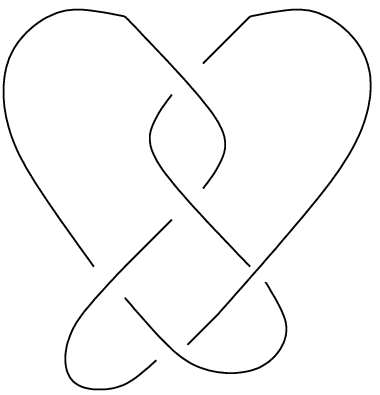}
\end{center}
\end{figure}
Consider a crossing of our knot observable as shown below,

\[ \epsfcenter{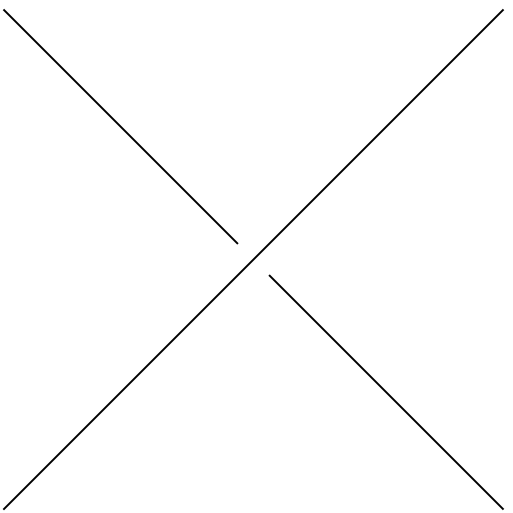} \]
Then to this crossing we assign a diagram as shown below

\[ \epsfcenter{heegard-1.eps} \longrightarrow \epsfcenter{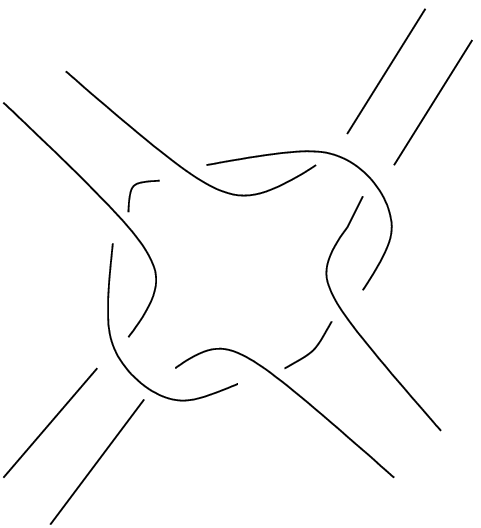} \]
The arrow just denotes this corresponding assignment.
Similarly for the opposite crossing. We can proceed with this for every crossing
of the knot or link, so that we will have a  diagram corresponding to a
knot.

\begin{figure}[h]
\begin{center}
\includegraphics[width=0.5\textwidth]{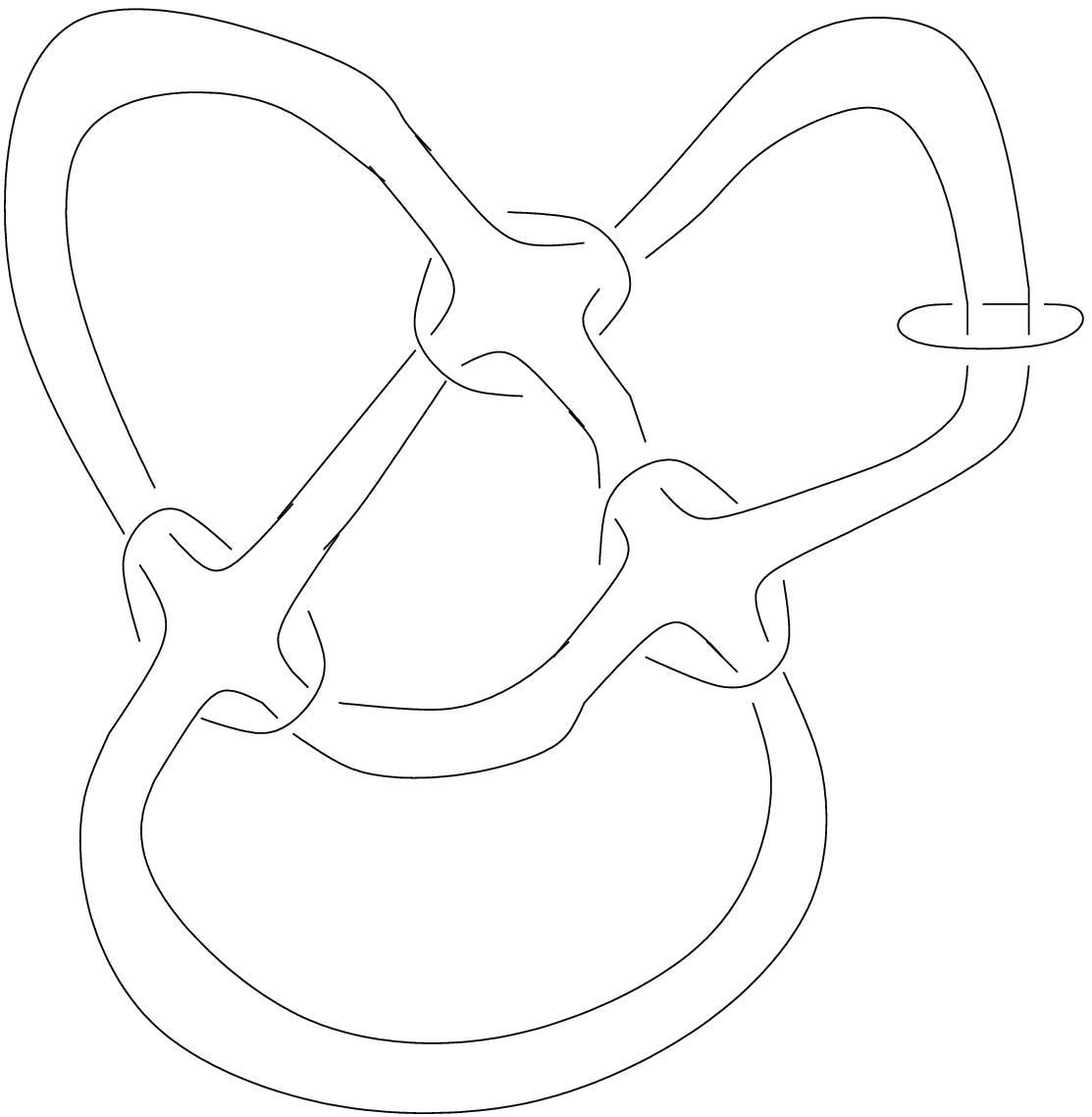}
\caption{}
\end{center}
\end{figure}
For example consider the trefoil knot and take its chain-mail diagram
as described.
Once all this has been done, we finally add one more circle around each 
component of the link, 
so that for the trefoil example, this looks as in figure 7.

We call these last circles special ones.
This diagram can be thought as a chain-mail diagram and in this way we can
define a way to calculate its expectation value.

\begin{definition}.
\danger{Chain-mail expectation value:}
Given the described chain-mail representation of our link observable, 
we define the "chain-mail expectation value" of it as follows:  attach 
$\omega = N^{-1/2} \sum_{j}^{r-2} \Delta_{j} \phi_{j}$ to each strand 
circle. To the special circles we attach a fixed representation 
$i \in \{ 0,...,(r-2)/2 \}$ of the quantum group. 
We then evaluate the value 
$CH(\mathcal{O})$ which is the chain mail value associated to
our observable link.\footnote{$\Delta_{i}=dim_{q}j$, and $\phi_{j}$
can be thought as the strand component coloured with the $j$ representation
(See the appendix, and also \cite{r}.}  
\end{definition}
This expectation value is a topological 
invariant of our observable. 
We then prove its invariance under the Reidemeister moves.
This tells us that we are in fact dealing with a knot and link
invariant.
  
\subsection{Invariance under Reidemeister moves}

We prove that the expectation value of our observables 
is invariant under the Reidemeister moves.
In order to prove the invariance under the first and second moves,
we simply draw the diagrams and use the known identities of
killing an omega, and two-strand fusion respectively. The
killing an omega identity tells us that when an $\omega$ circle
goes around an $\omega$ strand component, then the contribution 
is trivial. The two-strand fusion formula is just the special case
of the three strand fusion formula explained in the appendix when
one of the three strand components that go through the $\omega$ circle
is labelled by the trivial representation.

\danger{Invariance under the first Reidemeister move:}

\[ \epsfcenter{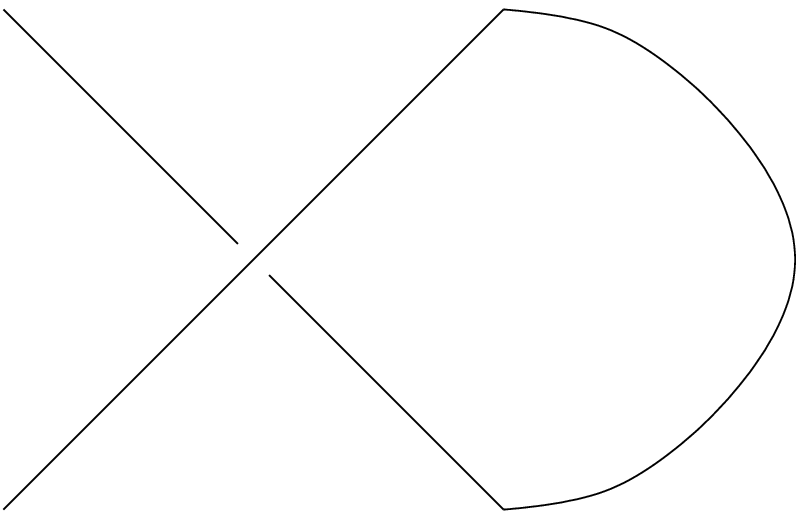} \qquad \longrightarrow 
\epsfcenter{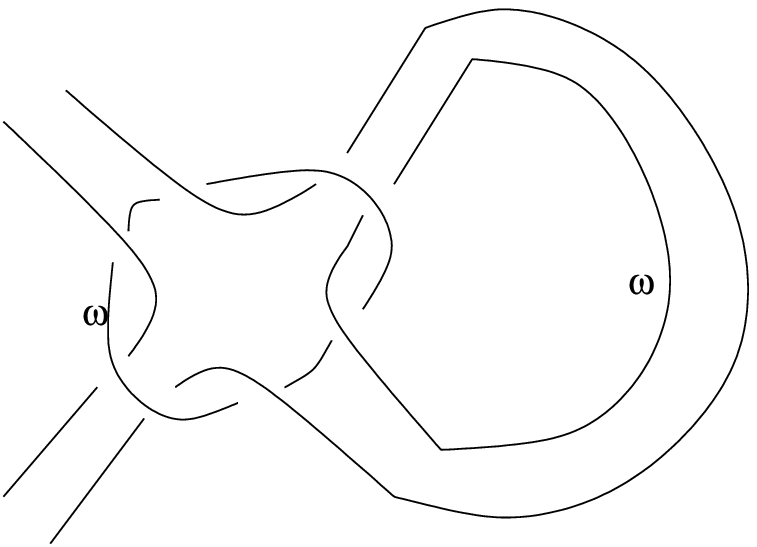} \]

\[ = \qquad
\epsfcenter{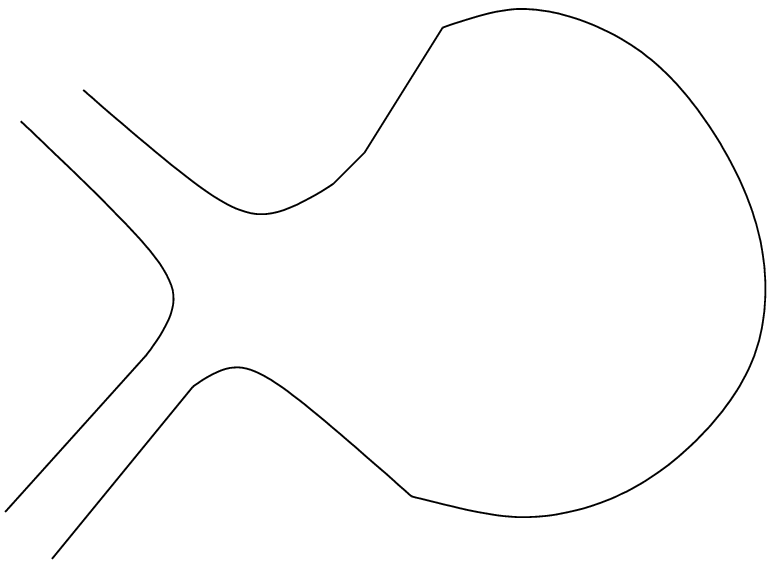} \qquad = \qquad \epsfcenter{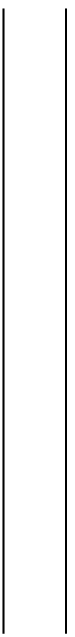} \qquad 
\longleftarrow \epsfcenter{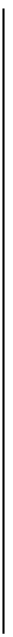} \]

\bigskip

\danger{Invariance under the second Reidemeister move:}

\[ \epsfcenter{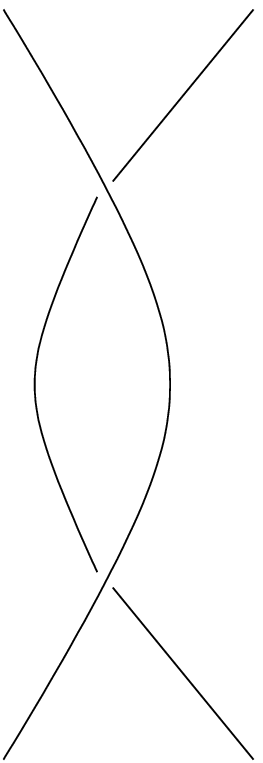} \qquad \longrightarrow \]

\[ \epsfcenter{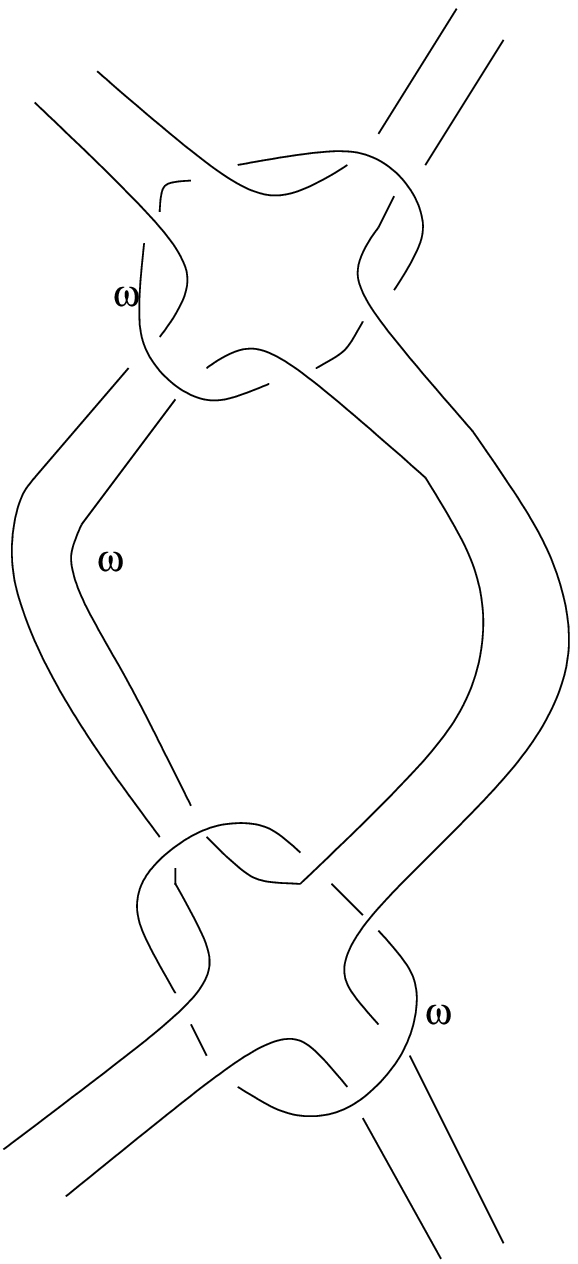} \qquad = \qquad
\epsfcenter{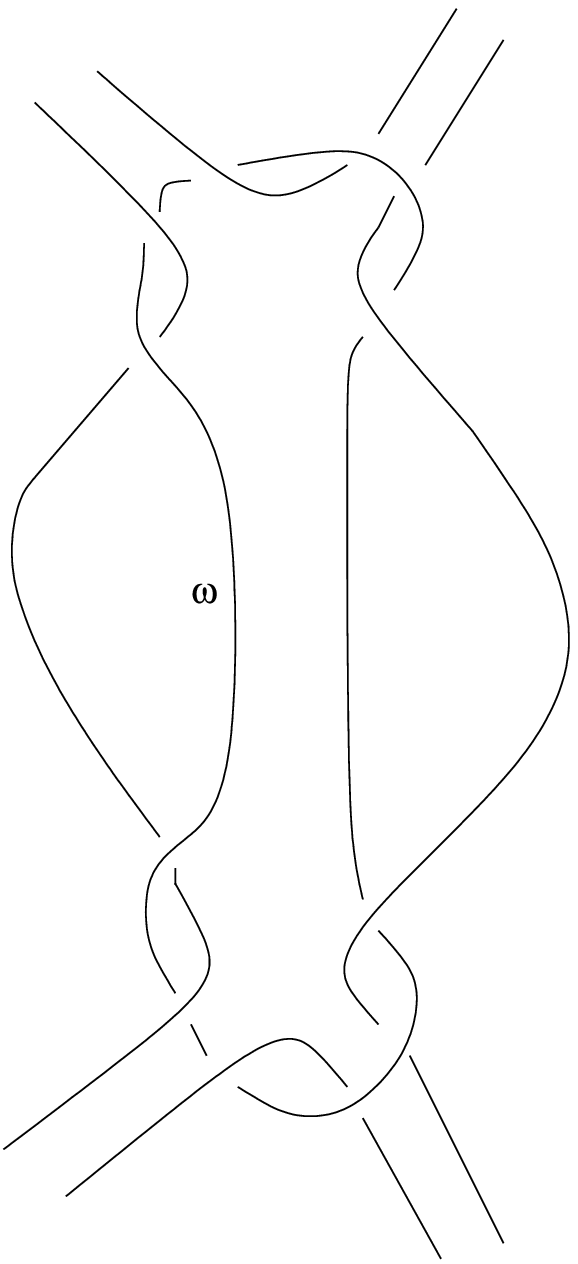} \qquad = \qquad \epsfcenter{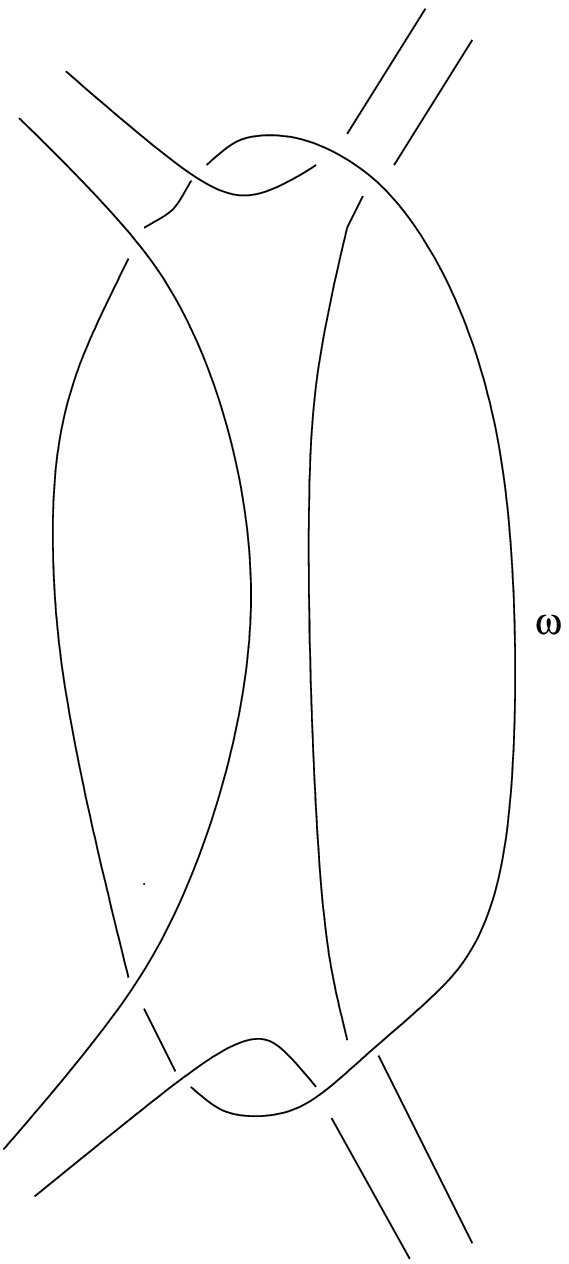} \]

\[ = \qquad
\epsfcenter{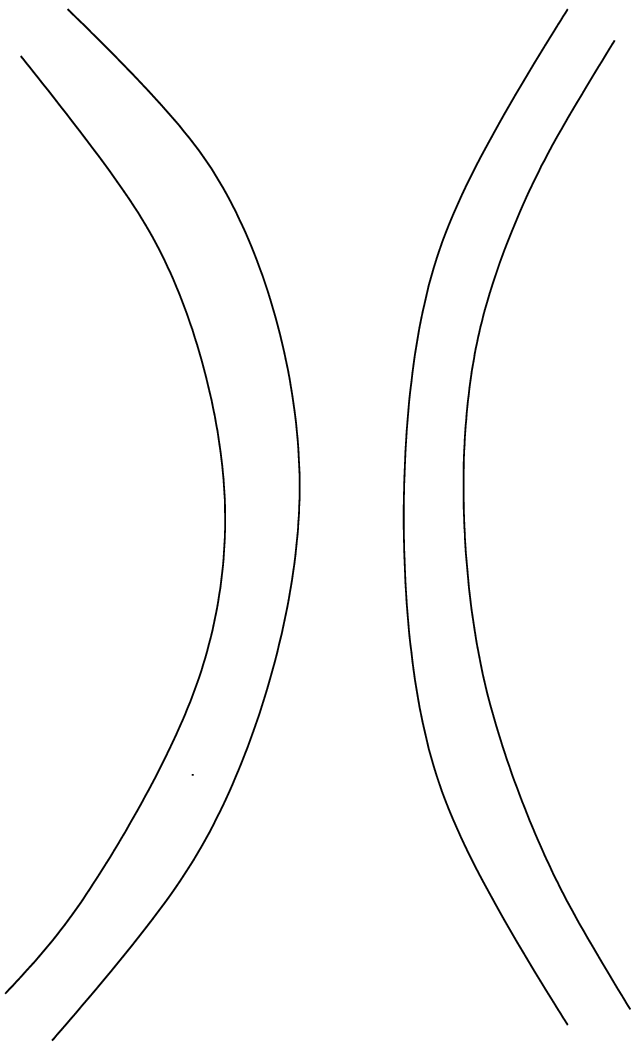} \qquad 
\longleftarrow \epsfcenter{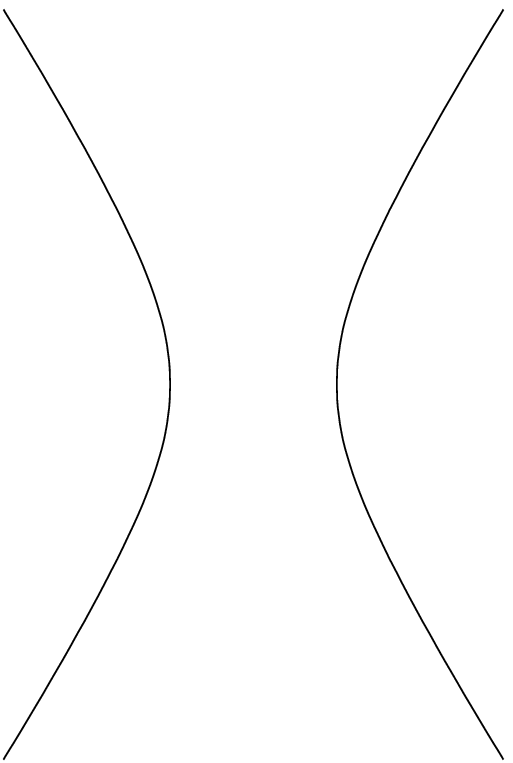} \]
Invariance under the third Reidemeister move cannot easily be computed 
by the above procedures.
To prove the invariance under this move requires the use of the 
formulas which will be developed in the following section.

\subsection{Computing the chain-mail expectation value}

We have already studied the chain-mail diagram corresponding to our knot or link
observable in $S^{3}$. Moreover, in the whole chain-mail diagram, we have
attached  $\omega = N^{-1/2} \sum_{j}^{r-2} \Delta_{j} \phi_{j}$ to all the components
except to the special  circles that go around the chain-mail diagram. 

Now, to compute what the expectation value might be for any knot or link
observable we use the following fusion strand formulas \cite{jm}:

\begin{equation}
\epsfcenter{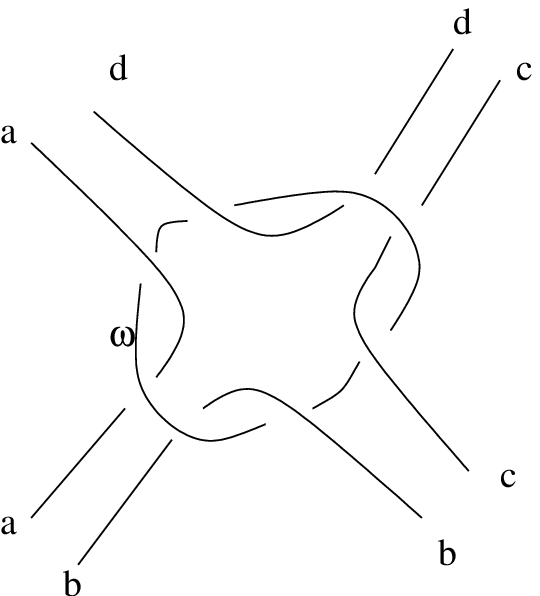}
 = N^{1/2} \sum_{i,j} \frac{\Delta_{i} \Delta_{j}}
{\theta_{abi} \theta_{cdi} \theta_{bcj} \theta_{adj}}\epsfcenter{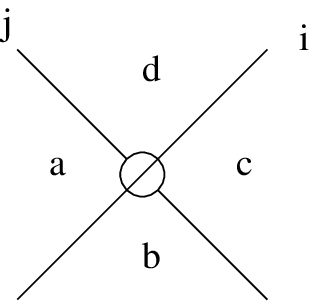}
 \epsfcenter{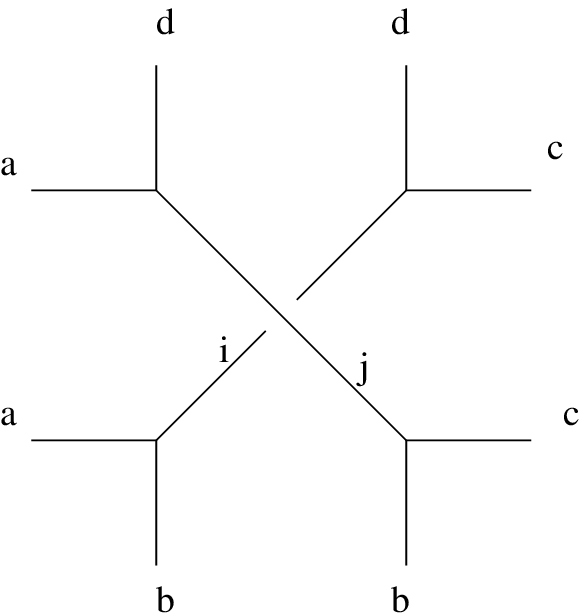} 
\end{equation}

\begin{equation}
\epsfcenter{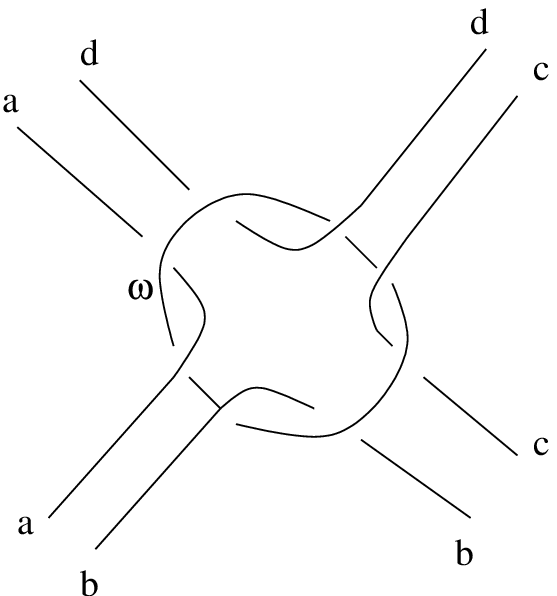}
 = N^{1/2} \sum_{i,j} \frac{\Delta_{i} \Delta_{j}}
{\theta_{abi} \theta_{cdi} \theta_{bcj} \theta_{adj}}
 \epsfcenter{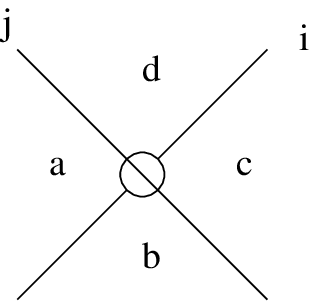} \epsfcenter{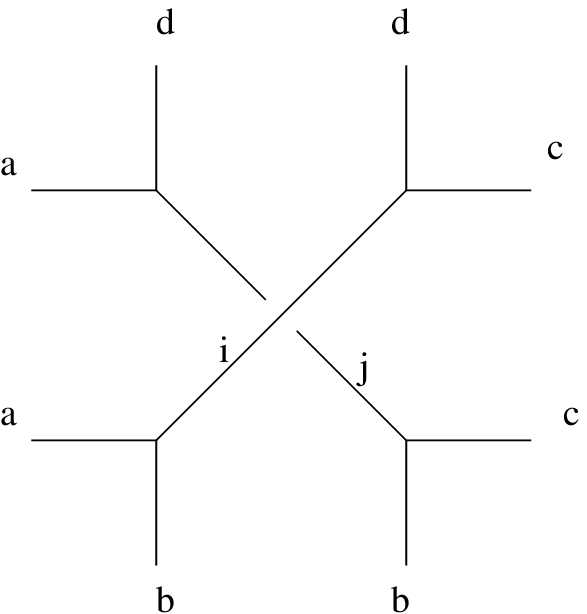}
\end{equation}
These formulas are useful to compute the chain-mail expectation value 
of any observable. Let us prove the invariance under
the third Reidemeister move.

\bigskip

\danger{Invariance under the third Reidemeister move:}

To prove invariance under this move, one associate to our diagram

\[ \epsfcenter{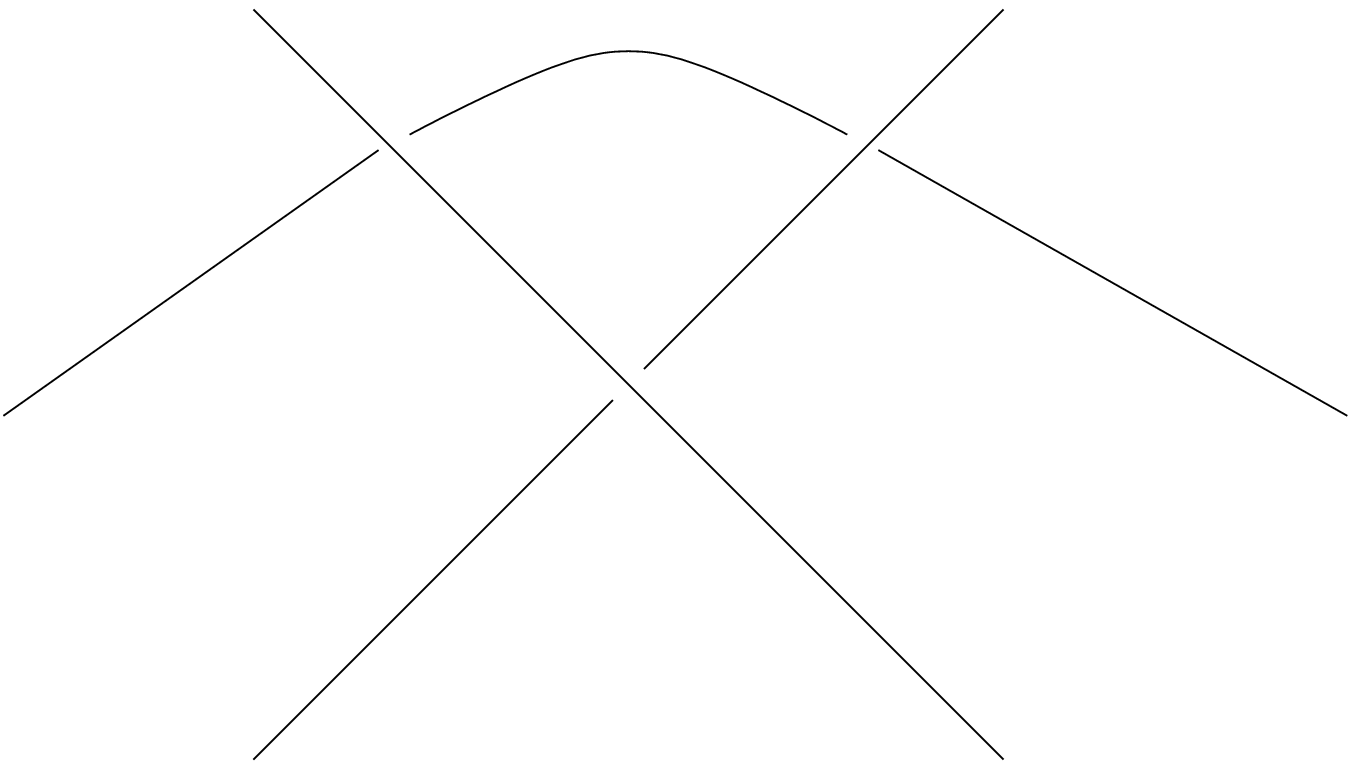} \]
the usual diagram

\[ \qquad \longrightarrow \epsfcenter{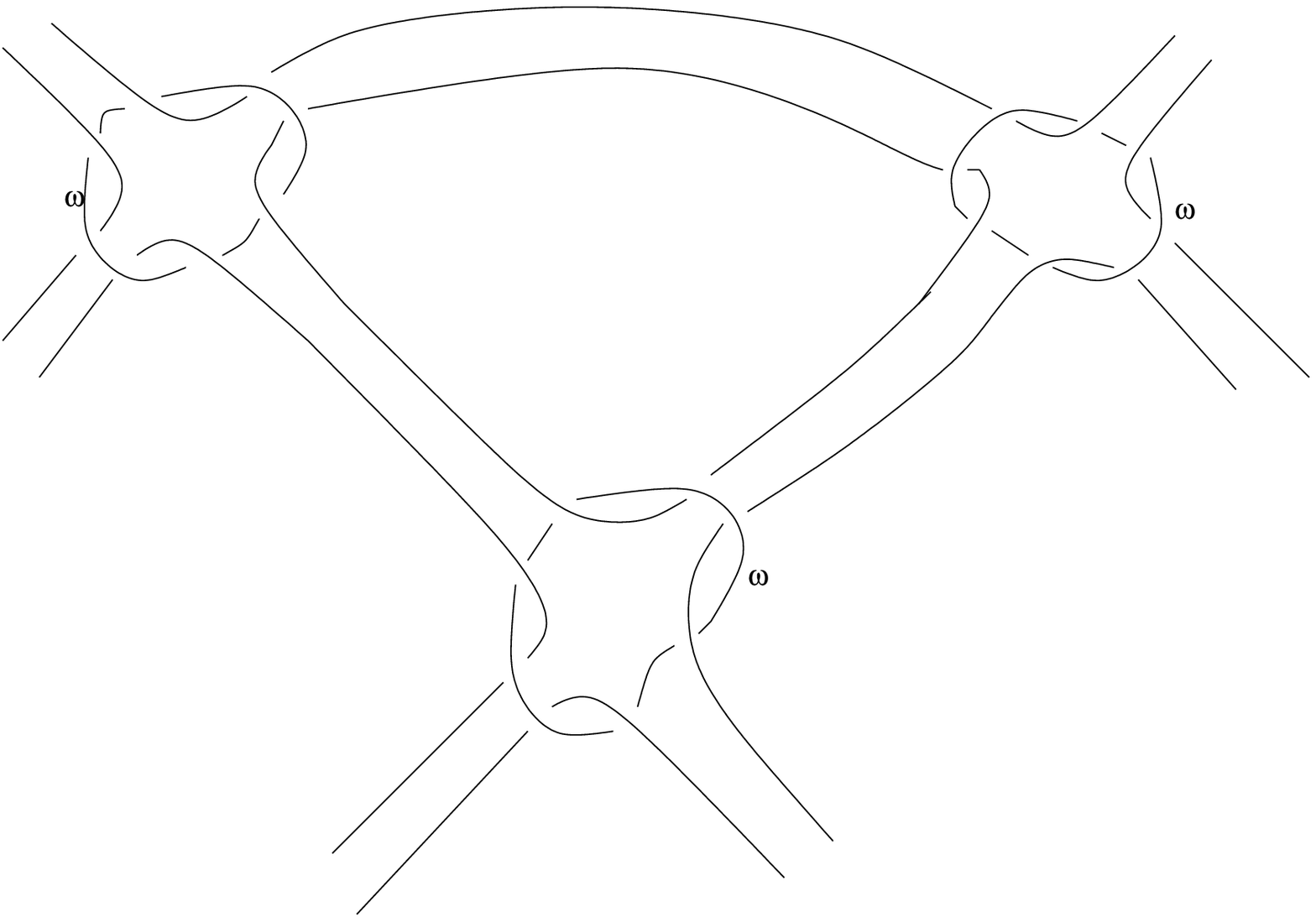} \]
We now apply equations (8) and (9) to the above diagram expanding it in a
sum and products of quantum dimensions, theta symbols, quantum 6j-symbols and 
crossing diagrams.  
If that is done also for the diagram

\[ \epsfcenter{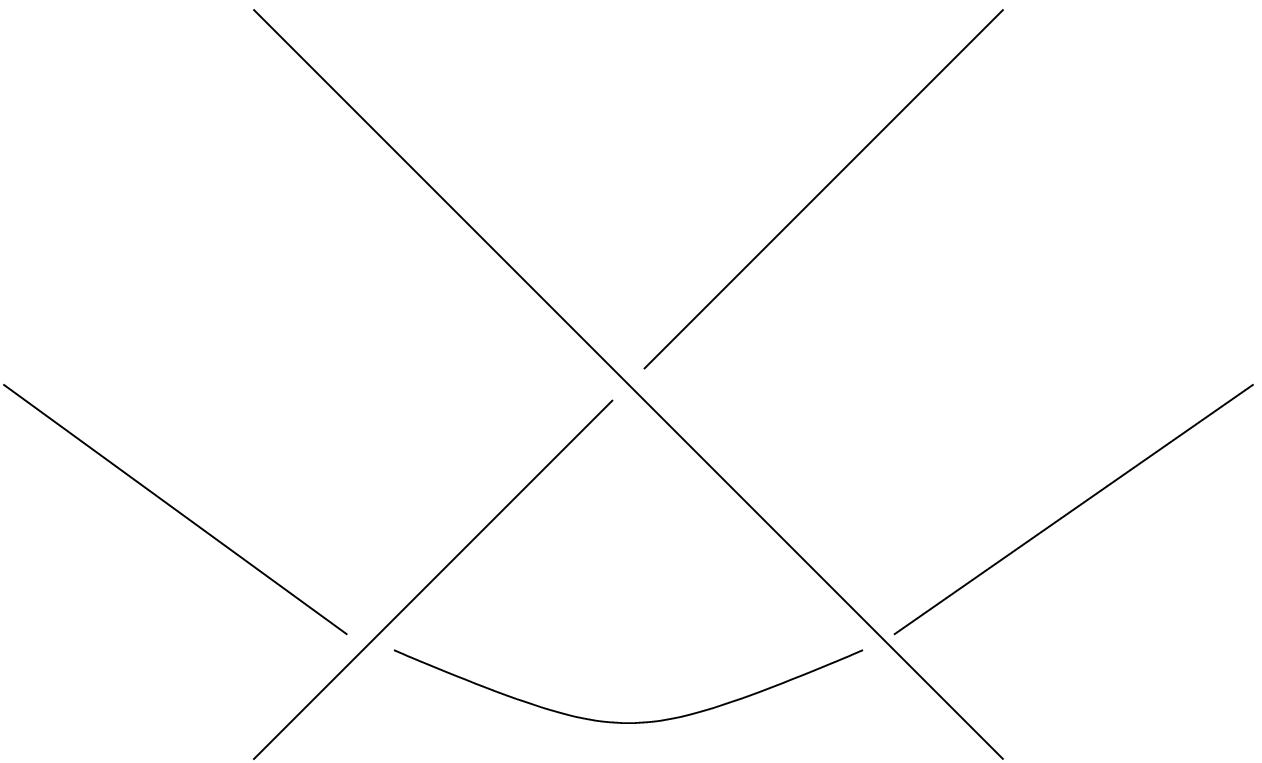} \]
we arrive at two formulas which are equal if 

\[ \epsfcenter{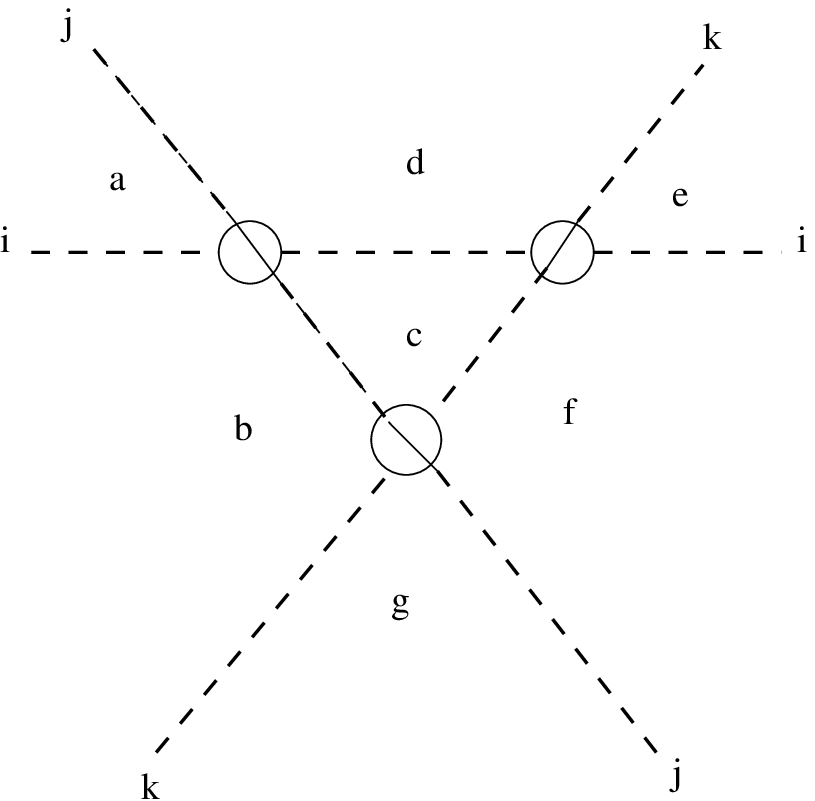} = \epsfcenter{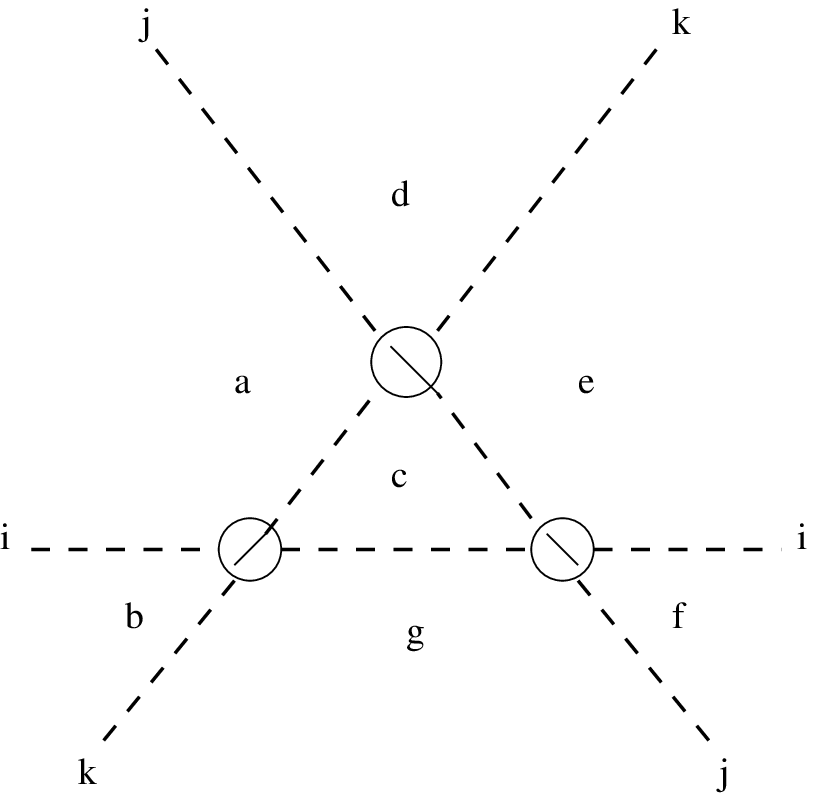} \]
but the above diagram formula is just the equality of two shadow world 
diagrams. (For a shadow world introduction we refer to \cite{kl} 
chapter 11)

The third Reidemeister move follows.

\subsection{Examples}

It is now a matter of calculation
to work out the chain-mail expectation value of any knot observale by 
following the
above technology. 
We present the result of the computation two examples.
The chain-mail expectation value of our first example was
computed step by step in \cite{jm}. 
The second one gives an interesting result where
we can see that there might be an
interrelation with Conformal Field Theory, as we will explain.

The examples we consider are the trefoil knot and the Hopf link.
 
\bigskip

\danger{The trefoil knot}

\bigskip

Consider the trefoil knot 

\[ \epsfbox{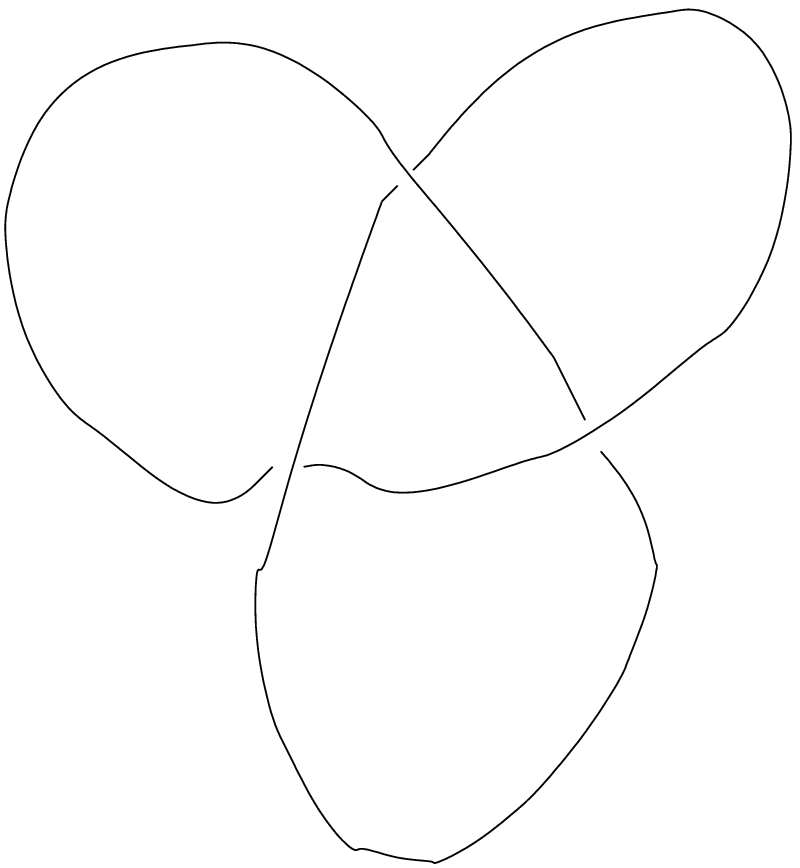} \]
Apply then the above construction to it as follows

\[ \epsfcenter{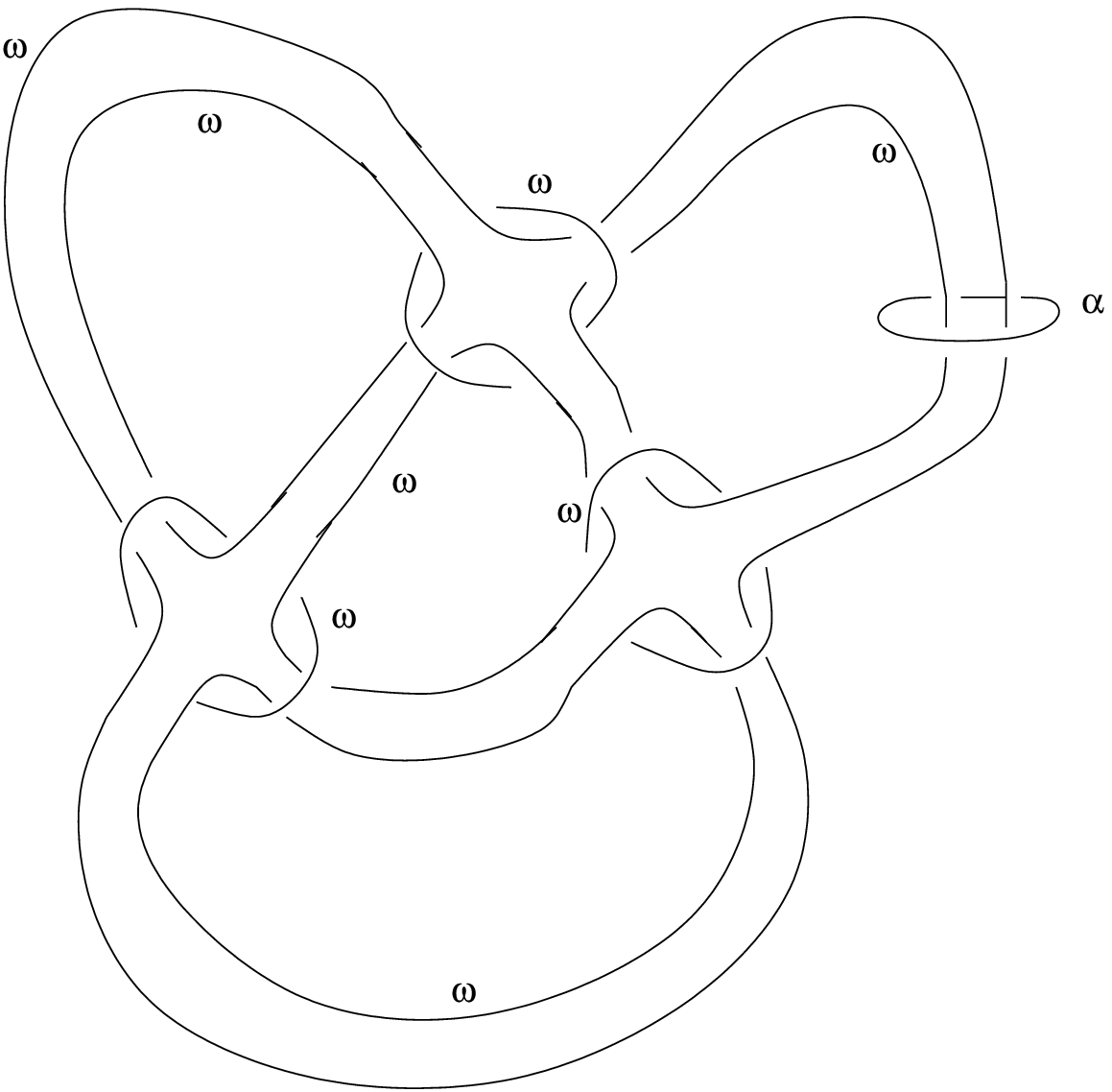} \]
then if we just apply the strand fusion formulas to this chain-mail construction
we show that for the trefoil knot 

\begin{equation}
CH(S^3, Trefoil)[\alpha] = N^{5/2} \sum_{i} \frac{S_{i \alpha}}{S_{i0}}
<Trefoil>_{R}  
\end{equation}
where $<Trefoil>_{R}$ means the relativistic evaluation given by 
the coloured Jones polynomial assigned to the trefoil knot 
times the coloured Jones polynomial assigned
to its mirror image \cite{b2}.

\bigskip

\danger{A general formula:} In fact it is easy to prove a simple formula which 
states that given a knot $K$
we have that its chain-mail expectation value is given by 

\begin{equation}
CH(S^3, K)[\alpha] = N^{(n/2)+1} \sum_{i} \frac{S_{i \alpha}}{S_{i0}}
<K>_{R}  
\end{equation}
where $n$ is the number of crossings of the knot, and $<K>_{R}$ is
its relativistic evaluation given by the coloured Jones polynomial
times the coloured Jones polynomial of its mirror image.

More generally, this formula extends to links as follows

\begin{equation}
CH(S^{3},L)[\alpha, \beta,...,\epsilon]= N^{(\# crossings /2)+1}\sum_{i,j,...k}
\frac{S_{i \alpha}}{S_{i0}}\frac{S_{j \beta}}{S_{j0}}...
\frac{S_{k \epsilon}}{S_{k0}}< L(i,j,...k) >_{R} 
\end{equation}
This general formula follows easily by observing that our fusion formulas
(8) and (9) are given by a product of a crossing diagram times a shadow world picture of the
opposite crosing of the link. The coefficients which appear in our
fusion formulas will be given by the shadow world formula of the complete
diagram of our knot or link. We will finally have a product of the 
coloured Jones polynomial of the knot or link times the coloured 
Jones polynomial of its mirror image. Finally, the special circles around 
each component of our knot or link give rise to the $S$ matrix factors.

\bigskip

\danger{The Hopf link}

\bigskip

This particular example is of great curiosity as it resembles some very well
known formulas of conformal field theory. In particular it makes us think
of whether there is a relationship between the chain-mail expectation value of our
observables and conformal field theory.

We do not know why this relation appears and it is interesting to search
for an explanation which might be hidden in a formulation in terms of
the $BF$ theory. Anyway, let us continue and  
consider the Hopf link

\[ \epsfcenter{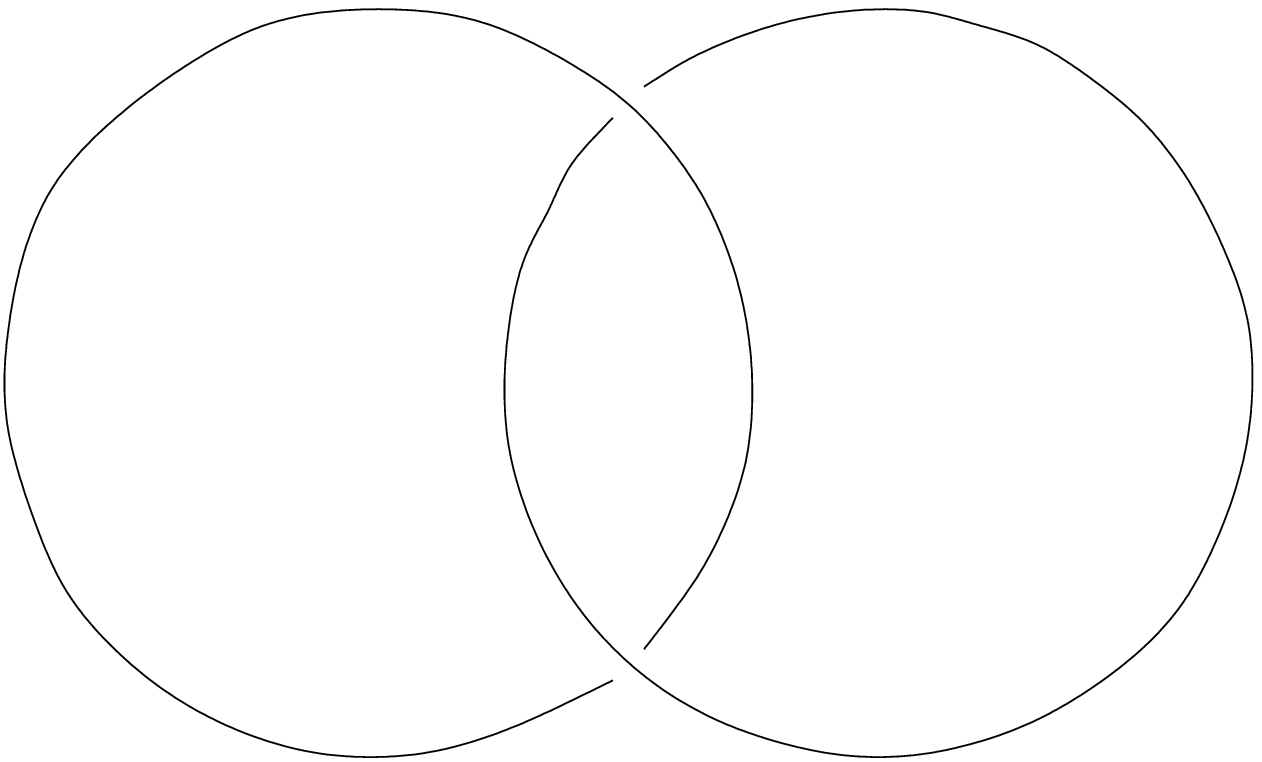} \]
and its diagram

\[ \epsfcenter{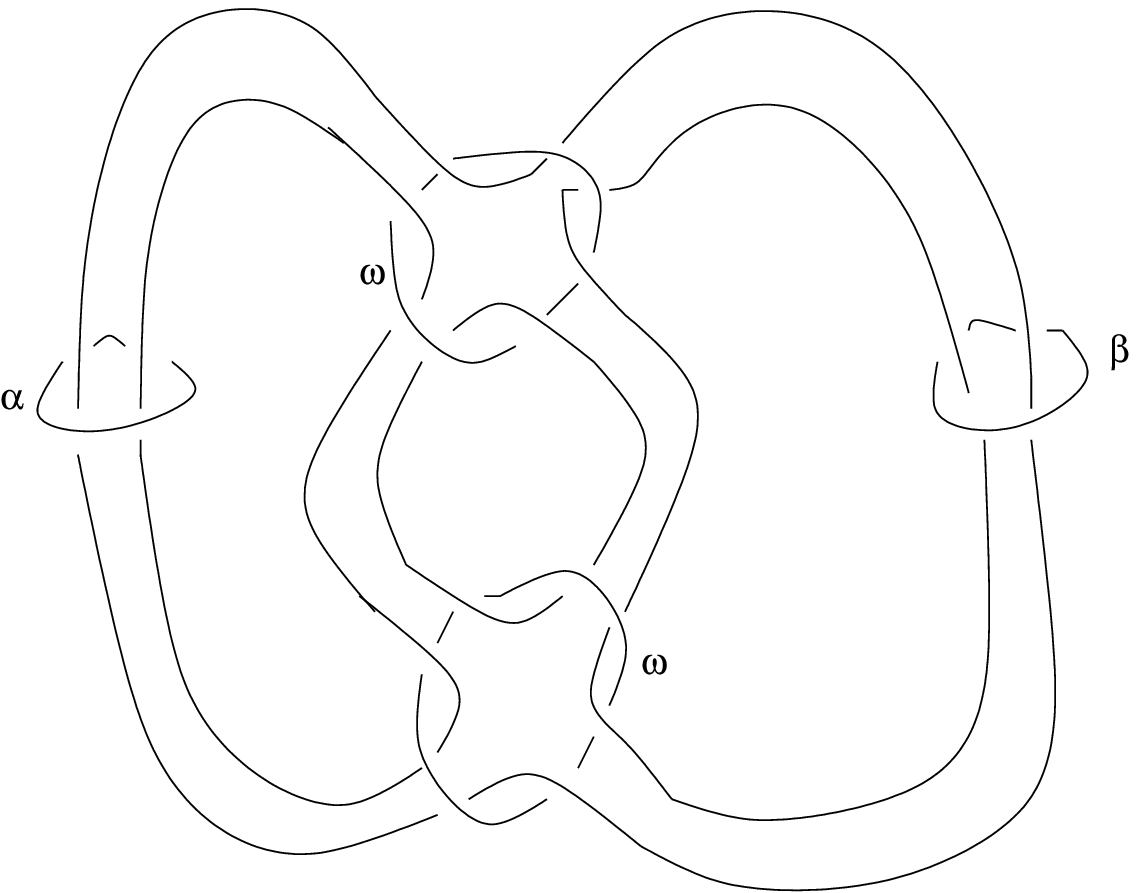} \]
Then its evaluation gives

\[ CH(S^3, \mathbf{H})[\alpha , \beta]= \sum_{i,j} \frac{S_{ij}}{S_{00}} 
\frac{S_{ij}}{S_{00}} \frac{S_{i \alpha}}{S_{i0}} \frac{S_{j \beta}}{S_{j0}} \]
Summing first over the index $i$ we have the relation 

\[ CH(S^3, \mathbf{H})[\alpha , \beta]= \frac{1}{S_{00}^{2}} \sum_{j} 
\frac{S_{j \beta}}{S_{j0}} \sum_{i} 
\frac{S_{ij} S_{ij} S_{i \alpha}}{S_{i0}} \]
If in the spirit of the Verlinde formula, we introduce the 
abbreviation $N_{\alpha j}^{j}$ for the sum over $i$ we obtain

\[ CH(S^3, \mathbf{H})[\alpha , \beta]= \frac{1}{S_{00}^{2}} \sum_{j} 
\frac{S_{j \beta}}{S_{j0}} N_{ \alpha j}^{j} \]

\[ \qquad \qquad = \frac{1}{S_{00}^{2}} \sum_{j} 
\frac{S_{j \alpha }}{S_{j0}} N_{ \beta j}^{j}   \]
If we had summed over the $j$ index first, we would had got a similar 
formula. Both formulas are the same and 
diagrammatically they may be written as

\begin{equation} 
\epsfcenter{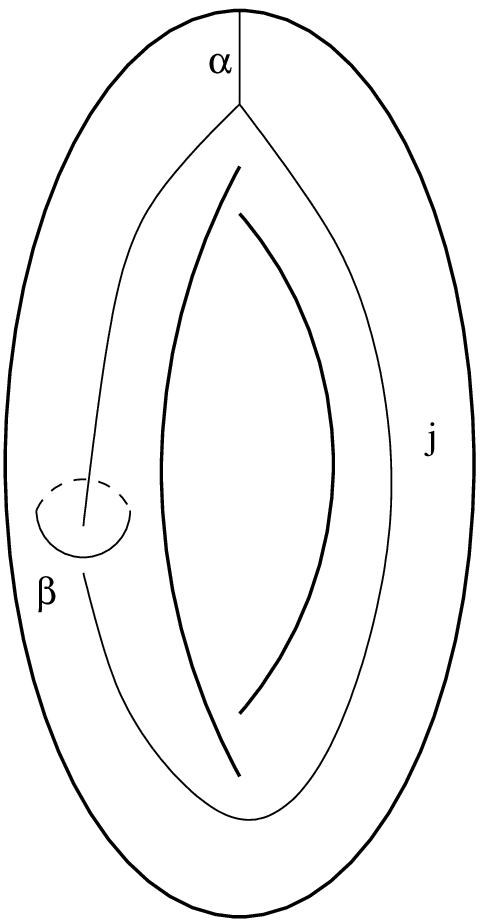} = \epsfcenter{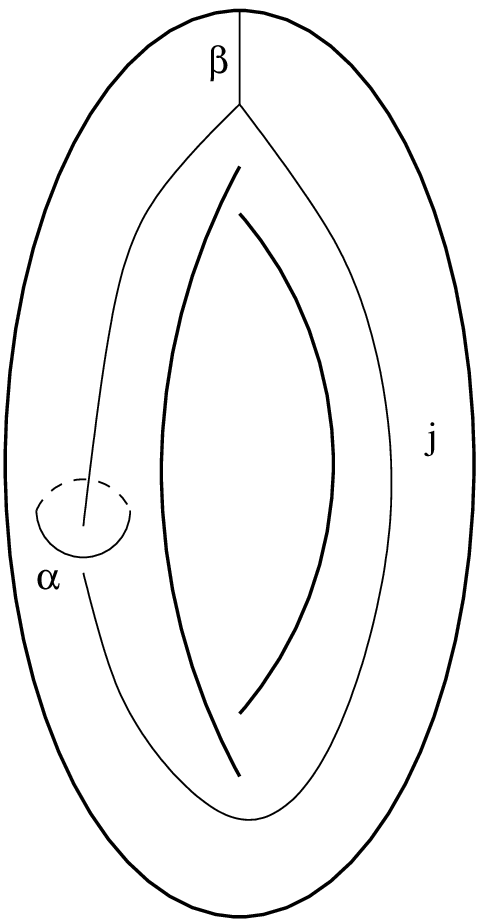} 
\end{equation}
This identity is an interesting example of an interrelationship 
between TQFT and conformal field theory, as it is a general case of two well
known formulas of conformal field theory. 

Let us say for instance that $\alpha$ or $\beta$ is trivial, 
then we have that 

\begin{eqnarray} 
CH(S^3, \mathbf{H})[\alpha]= \frac{1}{S_{00}^{2}} \sum_{j} N_{ \alpha j}^{j} 
\nonumber \\
\qquad \qquad = \frac{1}{S_{00}^{2}} \sum_{j} 
\frac{S_{j \alpha }}{S_{j0}} 
\end{eqnarray}
which diagrammatically can be expressed as

\[ \epsfcenter{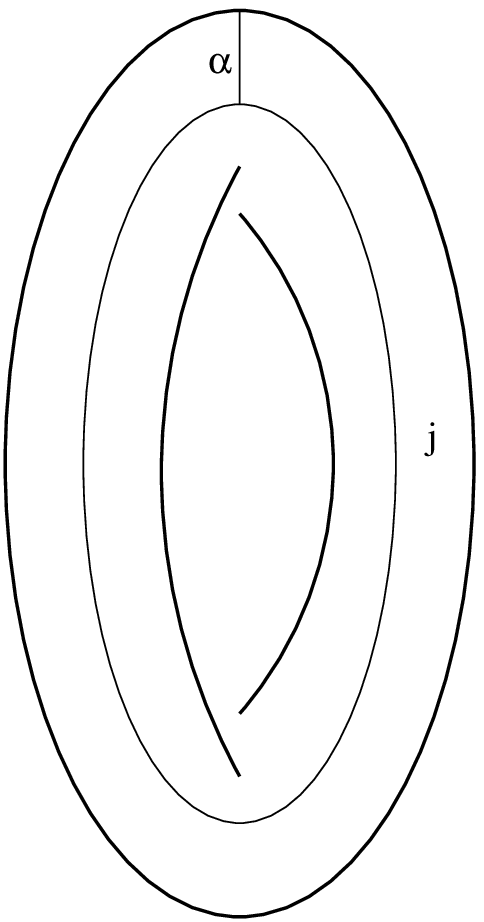} \]
This last formula is the dimension of the torus with one point charge
of conformal field theory.
\footnote{It is important to notice here that the dimension of the torus with
one point charge, and the dimension of the torus with no point charges
are to be understood in the context of conformal field theory,
and not to be confused with its dimension as a Riemann surface.}

Now, if both $\alpha$ and $\beta$ are trivial, then we have

\begin{equation} 
CH(S^3, \mathbf{H})= \frac{1}{S_{00}^{2}} \sum_{j} 
\delta_{j}^{j}= \frac{1}{S_{00}^{2}} dim \mathbf{T} 
\end{equation}
which is the dimension of the torus with no point charges on its boundary,
which is another well known fact of conformal field theory.

\section{Conclusions}

In this paper we presented a set of observables for 3-dimensional Riemannian 
quantum gravity with positive cosmological constant which also give topological
invariants of graphs embedded in 3-dimensional manifolds. For the case of
knots and links, we just dealt with the case of $S^{3}$.
Although the treatment of the topological invariance of the knots
and links observables embedded in any 3-dimensional manifold will
appear in \cite{b-jm}, a way to compute the expectation values 
by following an analogue procedure to the one we gave will be
interesting to describe. 

These observables have been described
in the discrete version of quantum gravity, and it is interesting
to find a description of these observables in terms of 
BF theory.

The treatment given here could shed some light in finding a set of
observables for more interesting
and physically relevant spin foam models. We can mention that at least
for other topological field theories there is a generalisation; 
for instance we can describe the same kind of observables for the Crane-Yetter model.

\newpage

\appendix{}

\section{Chain-Mail Invariants of 3-dimensional Manifolds}

In this appendix we introduce the chain-mail invariants of 
3-dimensional manifolds which were introduced by Roberts
\cite{r}.
This picture is a description of the 3-manifold as a special link formed 
by the attaching curves of the handle decomposition of $M$. The relation to the
Turaev-Viro model described in the previous chapter will be developed.
We consider again $M$ to be a closed, connected oriented 3-dimensional manifold
as in the previous chapter.

\begin{definition} 
Let $D$ be a handle decomposition of $M$ with $d_{0},
d_{1},d_{2},d_{3}$ handles of the corresponding dimensions. Let $H$ be the union
of the 0- and 1- handles, and $H^{\prime}$ be its handlebody complement, i.e. 
2- and 3-handles. Drawing the attaching curves of the 2-handles in $\partial H$ 
and then pushing them slightly into $H$ then adding the meridians of the
1-handles linking them locally in $H$ and finally giving framings to all these
curves, produces a kind of link which is called a $chain-mail$ $link$ denoted as
$C(M,D) \subseteq H$.
\end{definition}
An example of how it would look appears in figure 8

\begin{figure}[h]
\begin{center}
\includegraphics[width=0.5\textwidth]{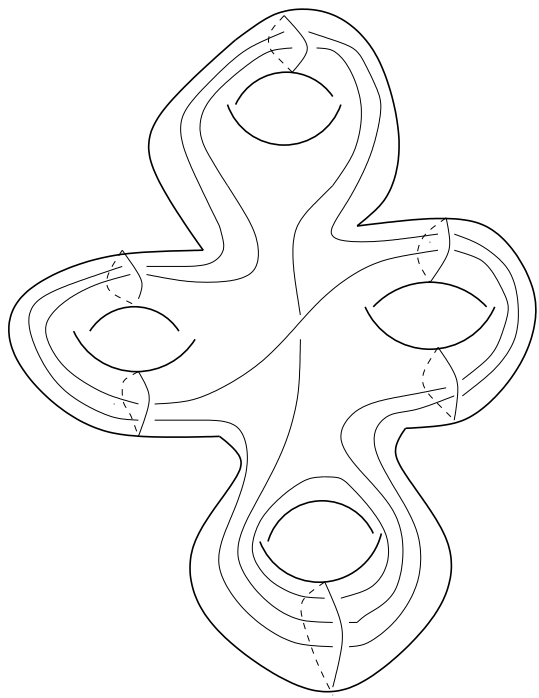}
\caption{Chain-Mail link}
\end{center}
\end{figure}
The next step is to embed our handle-body $H$ is $S^{3}$ so that we have a 
link. If we write $E$ for our embedding and $C(M,D,E)$ for the image of
$C(M,D)$ on $S^{3}$, we can now attach $\omega$ to all of its components, and
not forgetting about the framings, we obtain an element of $\mathbf{C}$ by
applying the fusion rules to it as before. Multiply now this element by
$N^{-d_{0}-d_{3}}$ and denote this last value as $CH(M,D,E)$.

In \cite{r} it was shown that the value $CH(M,D,E)$ is independent of the
embedding so we may just write $CH(M,D)$ for the above value.
Moreover, if $D_{1}, D_{2}$ are two handle decompositions of $M$ then
$CH(M, D_{1})=CH(M, D_{2})$.

Now we describe the relation between the chain-mail invariant and the
Turaev-Viro model described in the previous chapter.

Let $T$ be a triangulation of our 3-dimensional manifold $M$, and consider its
dual complex which is formed by placing a vertex inside each tetrahedron, and
one edge intersecting each triangle as shown for a tetrahedron in the 
figure below. We use arrows just to distinguish it from the other 
edges of the tetrahedron. 

\begin{figure}[h]
\begin{center}
\includegraphics[width=0.3\textwidth]{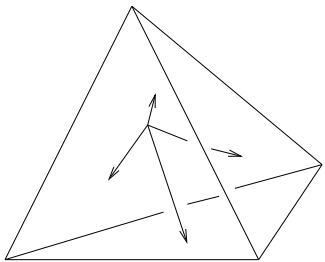}
\caption{Dual complex}
\end{center}
\end{figure}
The next step, is to thicken the dual complex of our triangulation and add 
curves $\delta_{j}$ corresponding to the face $f_{j}$, and curves 
$\epsilon_{i}$ corresponding to the edges $e_{i}$, so that we end up with a
chain mail $D^{*}$ which looks like figure B.3, for each tetrahedron.

\begin{figure}[h]
\begin{center}
\includegraphics[width=0.3\textwidth]{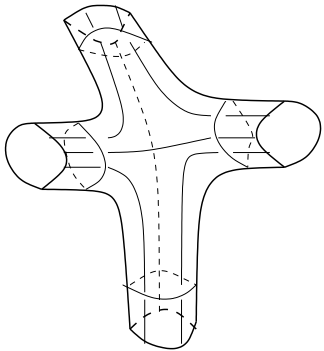}
\caption{Chain-mail picture of a tetrahedron}
\end{center}
\end{figure}
The relation between the Turaev-Viro partition fuction and the Chain-mail
picture is understood in the following theorem which is proved in \cite{r}.

\begin{theorem}
The chain-mail invariant of $M$ $CH(M,D^{*})$ equals the Turaev-Viro invariant
$Z(M)$.
\end{theorem}
The idea of the proof is to embed the chain-mail $D^{*}$ which we constructed 
from the triangulation into $S^{3}$ substituting $\omega$ along all the
attaching 2-handles and then use the 3-strand fusion formula defined by

\[ \epsfcenter{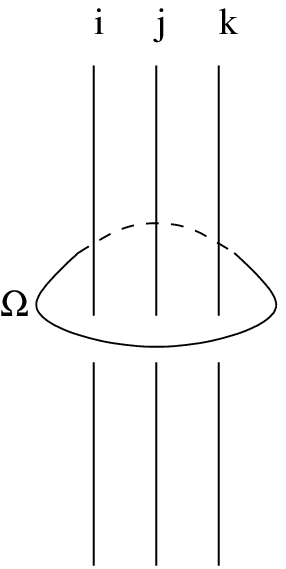} = \qquad \frac{N^{1/2}}{\theta_{ijk}} \qquad
\epsfcenter{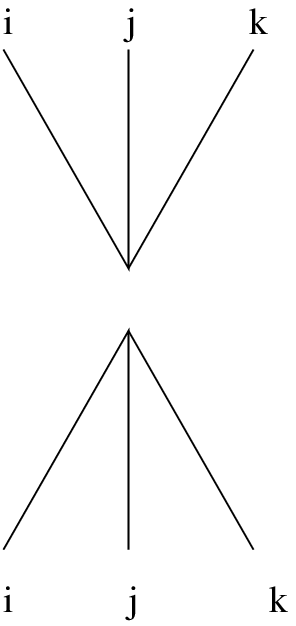} \] 
along all the $\delta$ curves(which correspond to faces of the triangulation).

The result will be a sum over all labellings of the 2-handles, of a product of
$\Delta_{i}$ coefficients associated to 2-handles, trihedron coefficients
associated to 1-handles, and tetrahedron coefficients associated to the
0-handles. This final sum after some careful observation equals the Turaev-Viro
state sum, obtaining then the equality $CH(M,T)=Z(M)$, and showing that the
Turaev-Viro partition function is equivalent to the Chain-Mail invariant of a
3-dimensional manifold. 

We are then in a position to use them interchangeably selecting the most
appropiate one to our needs.

\end{document}